# BLE Protocol in IoT Devices and Smart Wearable Devices: Security and Privacy Threats

Tushar Nagrare*, Parul Sindhwad, Faruk Kazi
*Electronics and Telecommunication Engineering Department, VJTI, Mumbai, India*
Email: *tpnagrare_m20@et.vjti.ac.in

***Abstract*—Bluetooth Low Energy (BLE) has become the primary transmission media due to its extremely low energy consumption, good network scope, and data transfer speed for the Internet of Things (IoT) and smart wearable devices. With the exponential boom of the Internet of Things (IoT) and the Bluetooth Low Energy (BLE) connection protocol, a requirement to discover defensive techniques to protect it with practical security analysis. Unfortunately, IoT-BLE is at risk of spoofing assaults where an attacker can pose as a gadget and provide its users a harmful information. Furthermore, due to the simplified strategy of this protocol, there were many security and privacy vulnerabilities. Justifying this quantitative security analysis with STRIDE Methodology change to create a framework to deal with protection issues for the IoT-BLE sensors. Therefore, providing probable attack scenarios for various exposures in this analysis, and offer mitigating strategies. In light of this authors performed STRIDE threat modeling to understand the attack surface for smart wearable devices supporting BLE. The study evaluates different exploitation scenarios Denial of Service (DoS), Elevation of privilege, Information disclosure, spoofing, Tampering, and repudiation on MI Band, One plus Band, Boat Storm smartwatch, and Fire Bolt Invincible.**

## I. Introduction

All IoT devices feature at least one sensor unit, enabling more direct integration between the natural environment and the computer system via communication protocols such as BLE, Wi-Fi, and ZigBee [1], [2]. BLE is an IoT communication protocol that focused on low power needs, fewer channel hopping, and improved security over prior versions. BLE, often known as Bluetooth Smart (BS), is the most widely used IoT communication technology [3]. It is a Wireless Personal Area Network (WPAN) era BLE initially included in the BCS (Bluetooth Core Specification) in June 2010 and has several superior features over regular bluetooth. IoT has been increasingly used in corporate systems, healthcare systems, army packages, beacons, novel household items, and various packages. There will be 15 billion connected IoT devices by 2022 [4]. Nearly all modern operating systems support Bluetooth and BLE, including Windows 10, Linux, Android, and Mac OS [5].

There are three steps to the BLE pairing process as shown in fig 2. First, both devices notify each other which pairing technique to employ and what the BLE device may do and anticipate in the beginning. A short-term key (STK) is generated and processed in stage two. In order to generate the STK, the two devices must agree on a temporary key (TK) that is combined with random integers. The STK is not transferred between devices. BLE is now utilized in billions of devices, it is critical to look at security flaws.

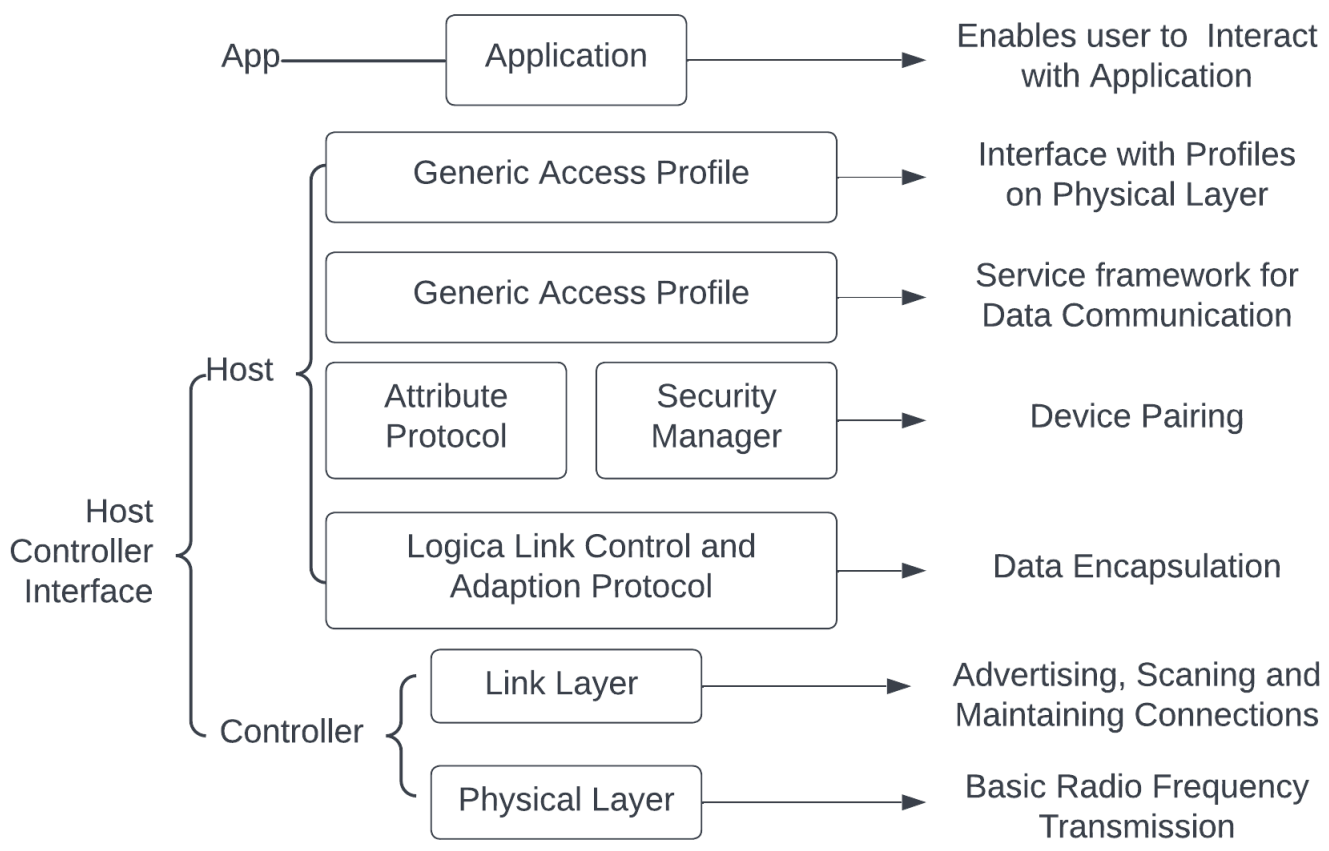

Fig. 1. BLE Protocol Stack [25]

Wearable devices like MI band 4, Boat Storm smartwatch, One-plus Band and Fire bolt Invincible smart watch these all personal and industrial gadgets make our lives easier but they also posses high threat. BLE's vulnerabilities might be fatal because of its broad use in healthcare applications. We found that wearable devices like MI band 4, Boat Storm smartwatch, One plus Band and Fire bolt Invincible smart watch has insecure pairing, inappropriate authentication, and poor protocol implementation exposes them to eavesdropping, pin cracking, and other attacks [6]. Security hazards include revealing personal information [7]. They write Personal security risks, identifiable information, and critical infrastructure attacks the Internet of Things developing technology is worrisome for instance, accounting and financial data storage [8]. He write Concerns around IoT trust and sensor integration include the secure IoT nodes are vulnerable because of threats to sensors[5]. Several studies on BLE security and privacy risks have been published independently, some researchers look at the protocol's security design and perform specific attacks that take advantage of the protocol's inadequate implementation [9], [10], [11]. Additionally, security researchers from academia and industry presented multiple attacks scenarios against IoT-BLE devices by performing attacks on IoT-BLE device via attack tools such as btlejack, adafruit, ubertooth one at various security conferences [12], [13]. This study aims to evaluate wearable device threat surface.

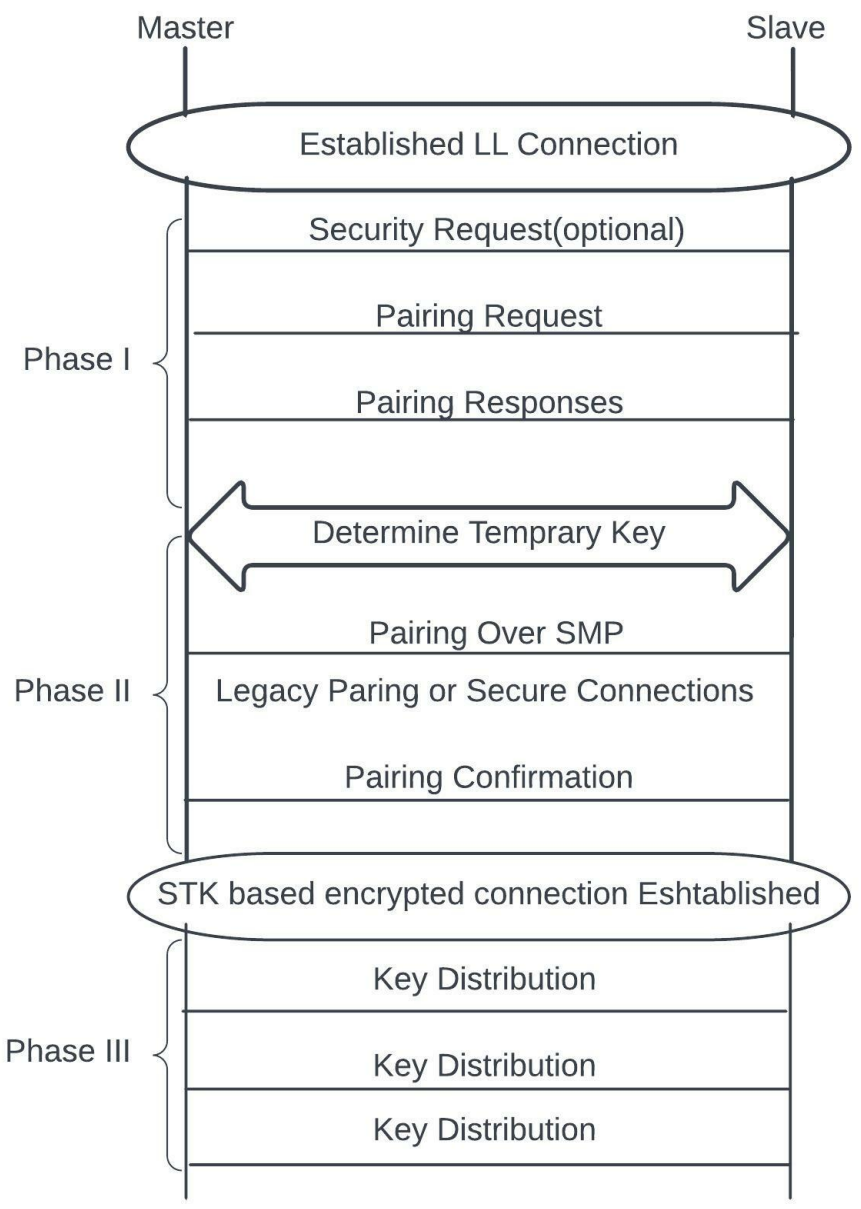

Fig. 2. BLE Connections Phases [3]

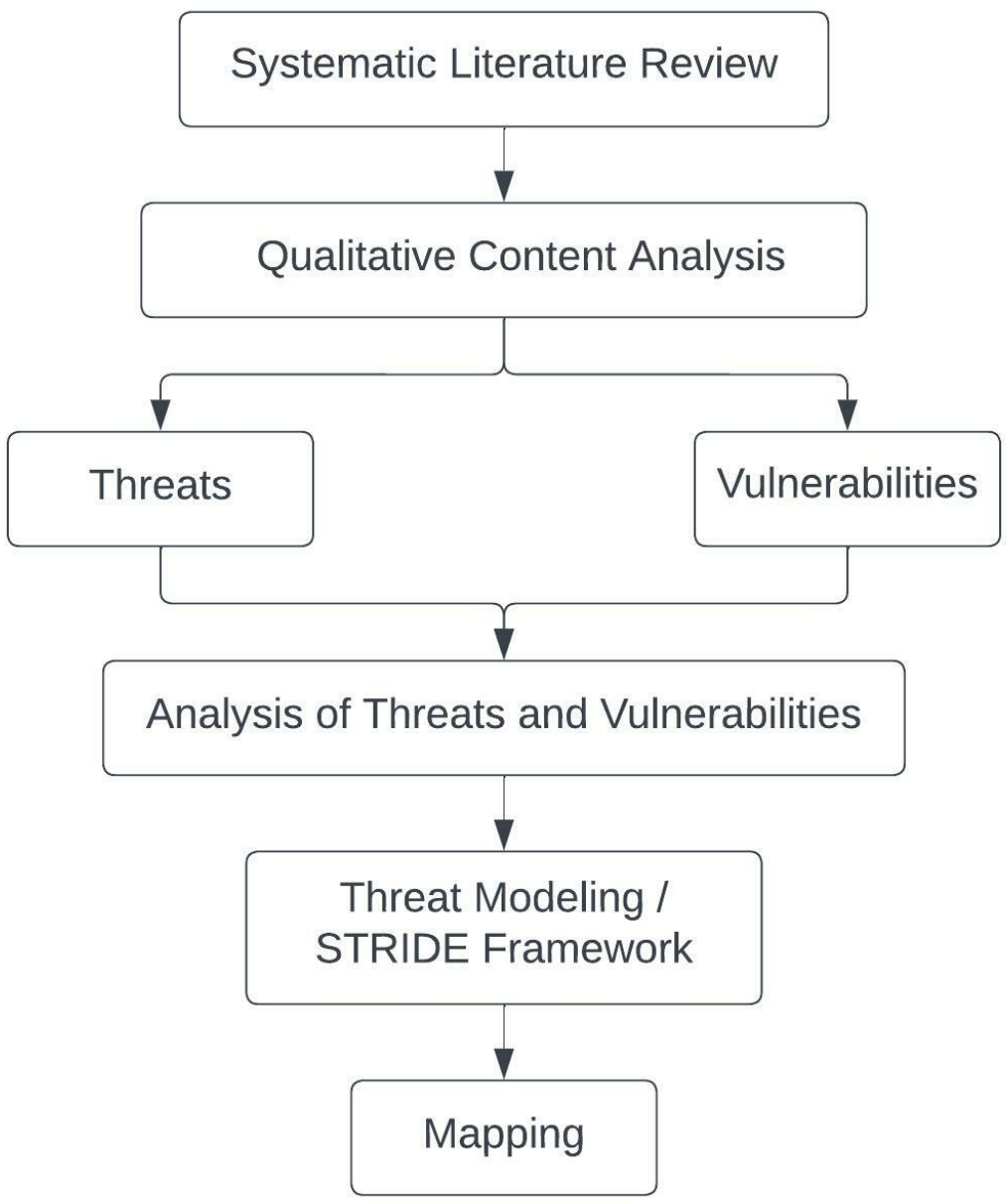

Fig. 3. BLE Analysis proposed model

This paper's contributions can be summarised as follows:

- With the help of STRIDE threat analysis of wearable devices, several vulnerabilities identified in multiple well-known (most recent) versions of the BLE protocol are utilized to categorize BLE risks.
- Present various examples of attacks on BLE wearable devices, along with the tools used to do them and discuss the lessons learned.
- Suggest mitigation technique for observed attack threat surface.

## II. Tools used for Attacking and Analysing BLE

This section details open-source security researchers that have created hardware and software tools to analyze various BLE vulnerabilities. Security researchers have used MITM, passive eavesdropping, and bluetooth encryption. These tools demonstrate security vulnerabilities on standard IoT devices as shown the analysis based in the mind map as shown. This list of open source, free tools help security researchers, BLE developers, and testers build up low-cost application components, identify and assess vulnerabilities, and safeguard BLE-enabled IoT devices. This section enlist various tools to perform attack on BLE wearable device.

### A. Hardware tools

*1) Ubertooth:* A free tool for exploring Bluetooth is called Ubertooth. Ubertooth offers gear that enables passive inspection of Bluetooth and BLE device communication. The January 2011-released Ubertooth One can detect and demodulate 2.4GHz Frequency band transmissions with a bandwidth of 1MHz [15]. Ubertooth sniffs the data and displays a visual of the local traffic sorted by frequencies and transmission intensity in dBm. This study approximates the overall Bluetooth traffic in the region, including distance diagnostics searching for items outside the purview of this project [4],[14]. The intensity signal strength on the channel frequency is used to structure the Ubertooth spectrum analyzer as shown in fig 12 the green lines show the strongest signal for that identified frequency, while the white lines show activities now. Wireless technology transmits data over the air, making it simple to intercept or tamper with these data packets.

*2) ADA-FRUIT BLE Sniffer:* Adafruit is a Bluetooth Low Energy (BLE) package capturing device similar to Ubertooth. Adafruit was created by Adafruit Industry[16]. It has the ability to intercept sent data packets. Analyze two BLE devices with this tool. Data gathered via Wireshark.

### B. Software tools

*1) Btlejack:* Btlejack is used to spy, jam, and hijack Bluetooth Low Energy devices. One or more of these devices are now supported, and one or more are running the special firmware. Based on the BBC Micro: Bit. This utility's latest version (2.0) supports BLE 4.0 and 5.0. However, only 1 Mbps uncoded PHY is supported by BLE 5.0 [14]. BtleJack controls the connection by actively disconnecting the master and moving it about in the connection, taking and sniffing.

*2) Wireshark:* Wireshark is a packet analyzer that is available for free. It is an open-source network protocol analyzer that frequently uses software for packet analysis. To help with data gathering, it may collect information via Ethernet, Bluetooth, USB ports, and other communication mediums Using a specific filter, Wireshark's graphical user interface (GUI) users can see captured packets in fig. 4 of some wearable devices as where vendors can't used encryption techniques data is transferred directly in plain text as seen.

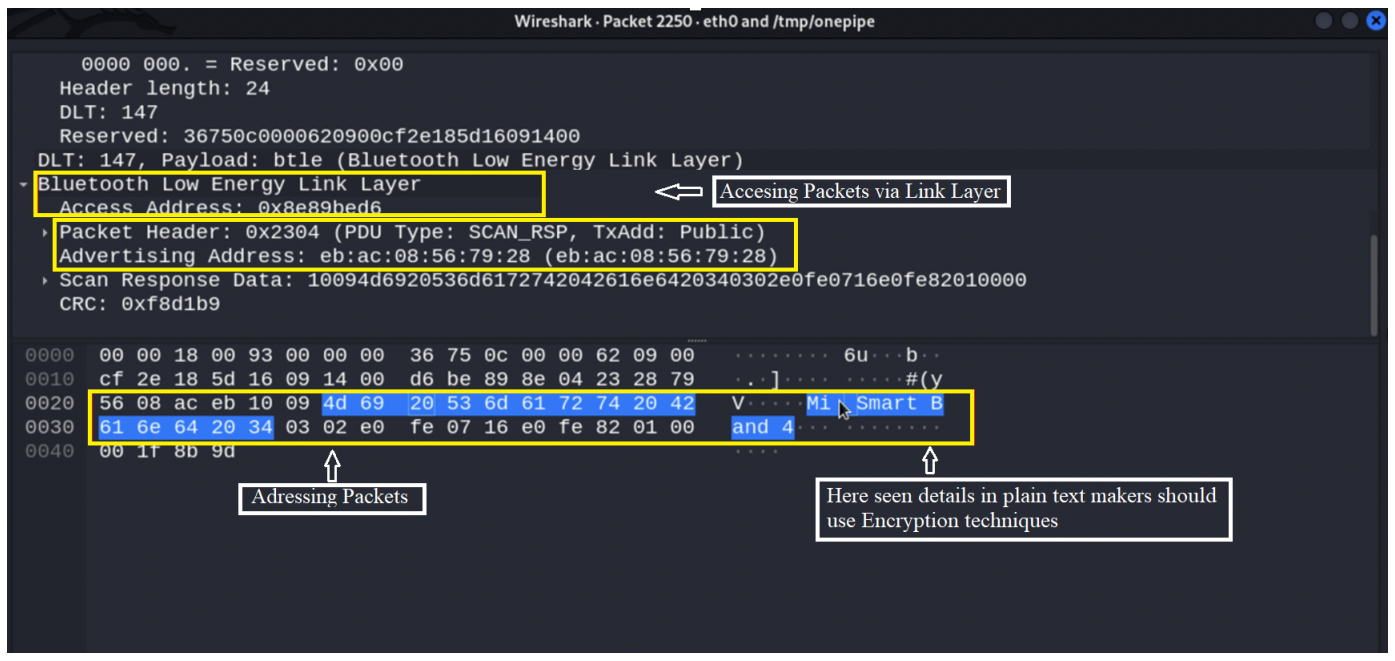

Fig. 4. Sniffed Packets in Wireshark

*3) Btlejuice:* Cauquil created BtleJuice, which he exhibited at the 2016 DefCon24 conference as a MITM attack against BLE devices. Intercept score, intercept proxy, specialized web interface, python, and Node JS binding are the four components of Btlejuice. The key components are interception cores and proxies. These two components function independently and communicate via the Web Socket protocol btlejuice functions as a proxy between BLE accessories and mobile applications.

## III. Security Prerequisites from STRIDE Threat Analysis

Looking at the DFD of wearable devices, we can derive security requirements and evaluate the attack scenarios acquired from the analysis of wearable devices like MI band 4, Boat Storm smartwatch, One plus Band, and Firebolt Invincible smartwatch since, Determining security needs, we consider the attack scenarios derived from the DFD diagram since the security requirement is meant to stop attackers from attaining their objectives, below are some security requirements.

*1) Adding timestamp in the packet:* Incorporating a timestamp into the transmission for all data exchange and connection processes, a timestamp should be included in the packet to avoid replay attacks.

*2) Authentication:* Even if an attacker edits a message, the message authentication mechanism can identify the change and prevent it from being inserted. Also, to avoid connecting to an attacker's device, stealing accounts, and circumventing the login procedure, the user authentication process should be employed.

*3) Encryption:* An attacker's ability to read and change the original communication is hampered by encryption. As a result, message alteration and network packet sniffing should be prevented by the packet encryption procedure. Additionally, before saving any sensitive data in the database, it must be encrypted. The attacker won't be able to access the data with even physical access to the storage.

*4) Traffic analysis and intrusion detection:* Attack detection and traffic analysis It's tough to stop excessive data transmission completely. As a result, we must mitigate it on the server side using traffic analysis and penetration testing.

*5) Secure account management:* Account management that is safe to avoid account theft, all users must maintain their accounts securely. The service provider should inform the user about the possibility of account theft and safeguard the data store.

The Microsoft threat modeling tool provides a threat modeling report based on our model as data flow diagram shown in fig 5. It generates 111 threats on its own. Based on prior research on smart band security,and identified 44 significant threats that may achieve the attacker's aim. also, given the practical analysis below in attack analysis section A and B with all attack scenario on IoT-BLE wearable devices.

## IV. Attack Analysis of Wearable Device

The security architecture of BLE is distinct from that of classic bluetooth. Low power consumption, computationally restricted sensors, and connection with IoT devices are all supported by BLE. With BLE, we have to choose between performance, security, privacy concerns, and low power usage. Different ways for secure connections are included in the BLE standard. Device binding, link-layer encryption, and device safe-list Unfortunately, many IoT-BLE devices do not implement these security methods properly. As a result, numerous security threats arise because many of these risks are caused by vulnerabilities in a common underlying architecture or protocol, the same mitigating procedure applies. Some attacks are mutually beneficial. This section focuses on how to arrange and perform assaults using the threat model sequence.

The user information is sent between the smartphone and the smart band as demonstrated in the DFD in fig. 5 smartphone. also shown each threats to the specific BLE wearable devices (MI Band 4, Boat Storm Smart watch, Oneplus Band and Fire bolt invincible) as shown in the table identified threat for analysed wearable devices which derived from the STRIDE threat analysis. Examining the system with an emphasis on user data and the smart band service should be done to overcome the security issues in BLE devices.

### A. Passive Attack

The first step in any attack scenarios is passive listening. Assailants secretly monitor every communication between linked devices, resulting in a wide range of destructive attacks. By interfering with data transmission somehow, we can eavesdrop on and intercept every conveyed data. Passive eavesdropping attacks are particularly vulnerable since data is delivered wirelessly, and an attacker only needs an interceptor such as Micro bit V2 to intercept wireless communications of devices

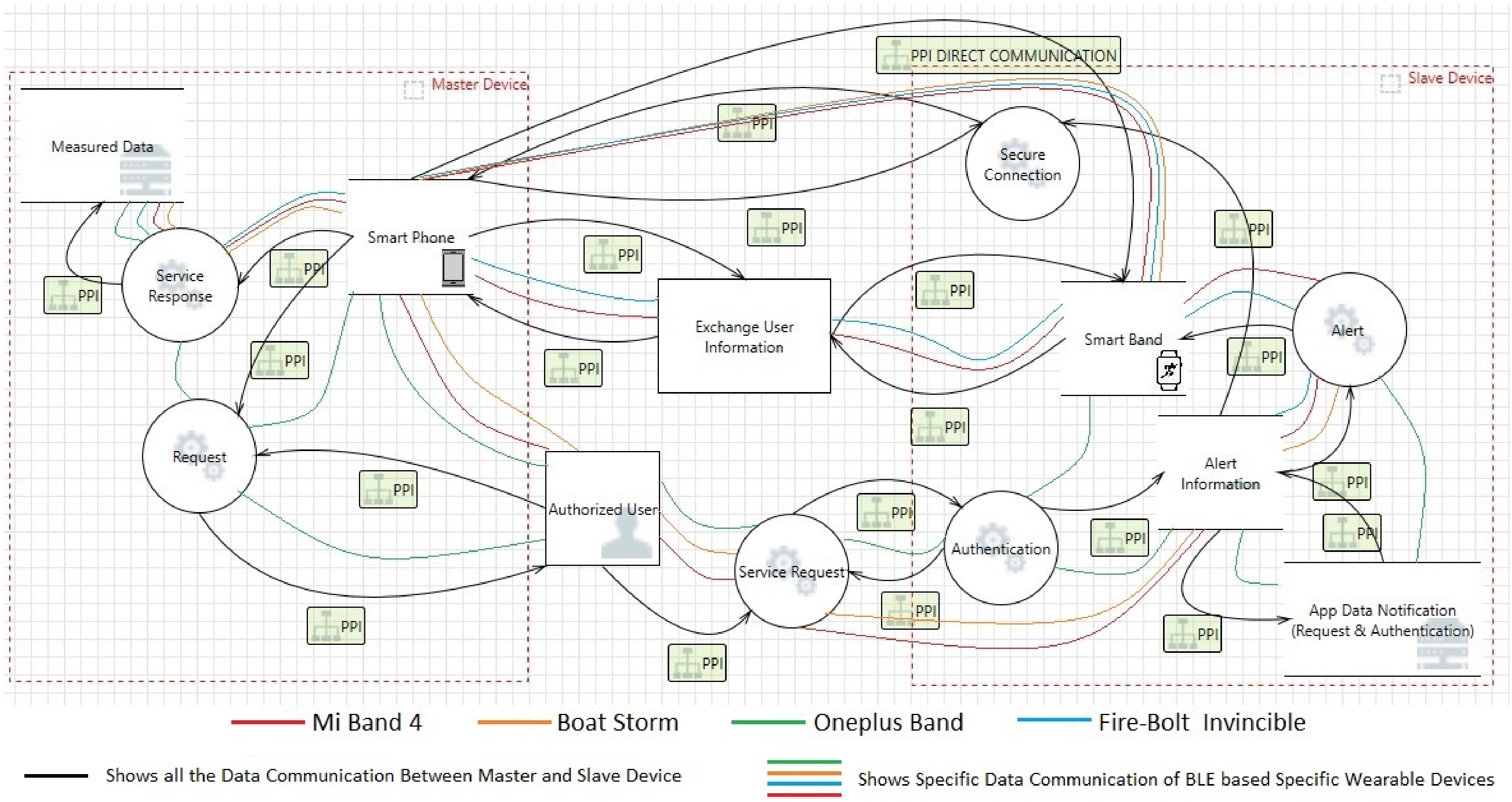

Fig. 5. Data Flow Diagram

TABLE I
IDENTIFIED THREATS FOR ANALYSED WEARABLE DEVICES

| | Denial of Services | Elevation of Privilege | Information Disclosure | Repudiation | Spoofing | Tampering |
|---|---|---|---|---|---|---|
| **Mi Band 4** | Yes | No | Yes | No | Yes | No |
| **Boat Storm Smart Watch** | No | No | Yes | No | No | No |
| **One Plus Band** | Yes | No | No | No | Yes | No |
| **Fire-Bolt Invincible** | Yes | Yes | Yes | No | Yes | Yes |

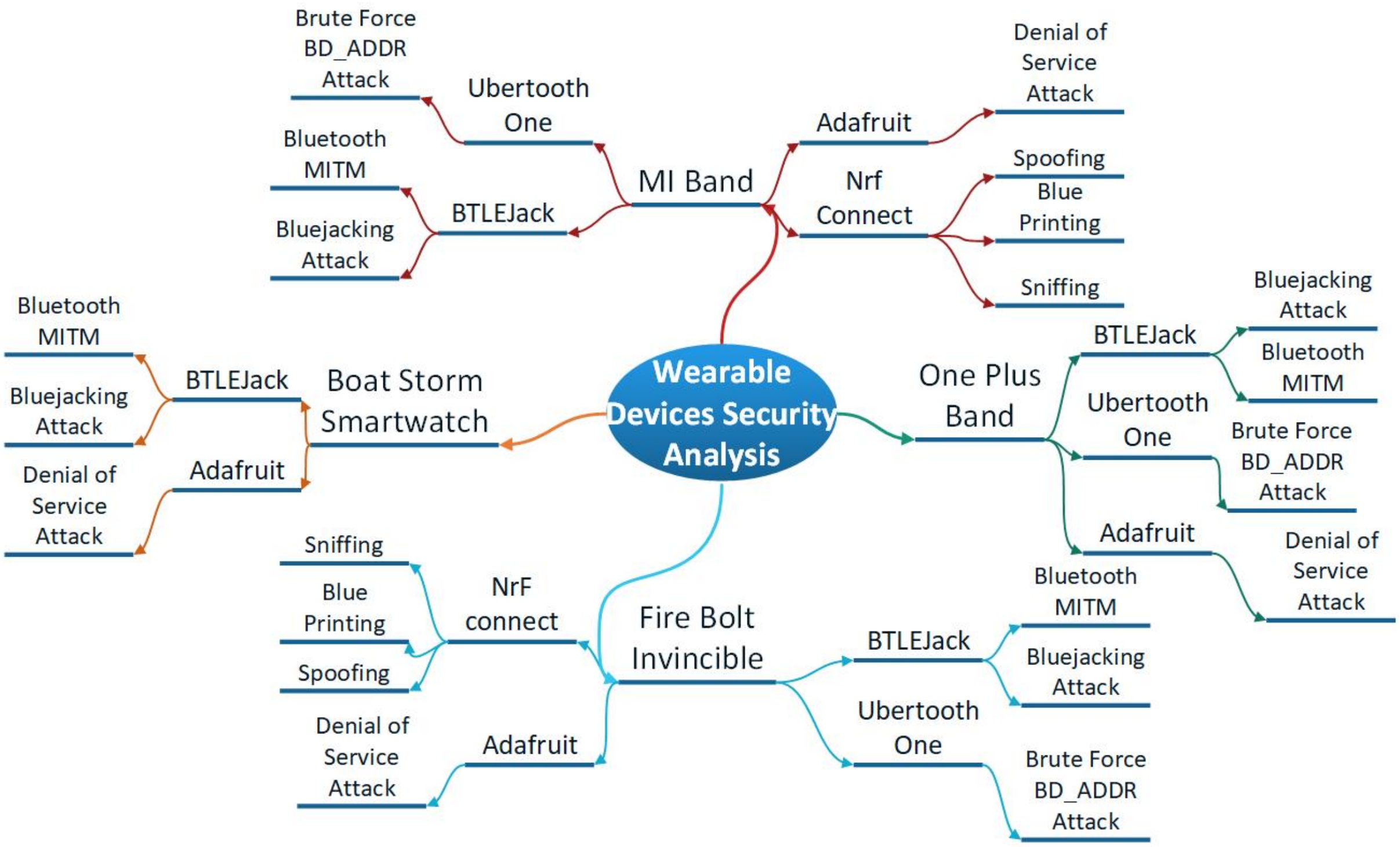

Fig. 6. Security Analysis Mind-map

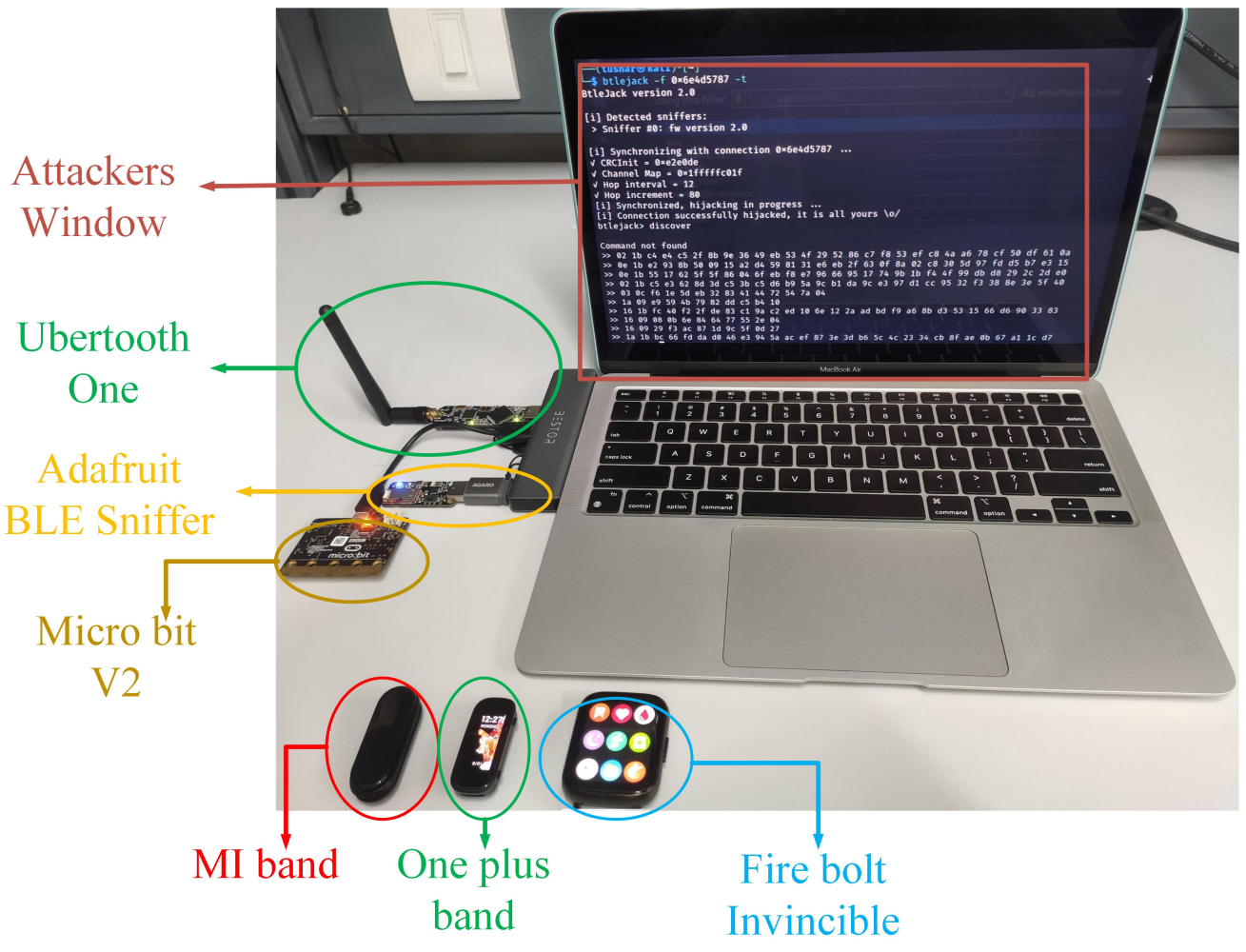

Fig. 7. Attackers Testbed

(MI Band 4, One plus Band and Fire bolt invincible) this IoT-BLE wearables are particularly vulnerable to this attack due to its predictable and obvious channel switching.

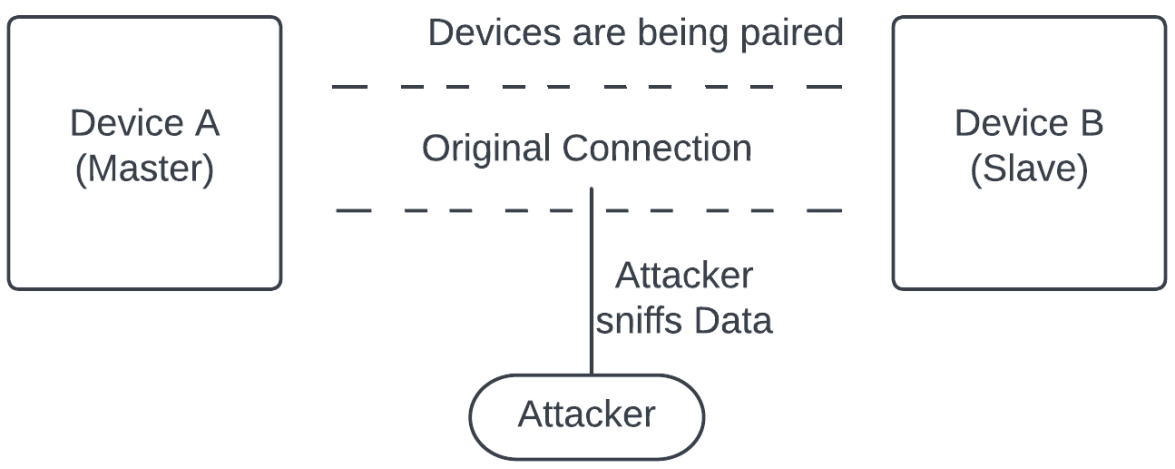

Fig. 8. Passive Attack

Most BLE systems like MI Band 4 and fire bolt invincible smart watch and Boat storm are vulnerable to this sort of attack due to the insufficient protection of the BLE standard and poor encryption algorithms, as well as several critical exchange mechanisms attackers to decode data easily.

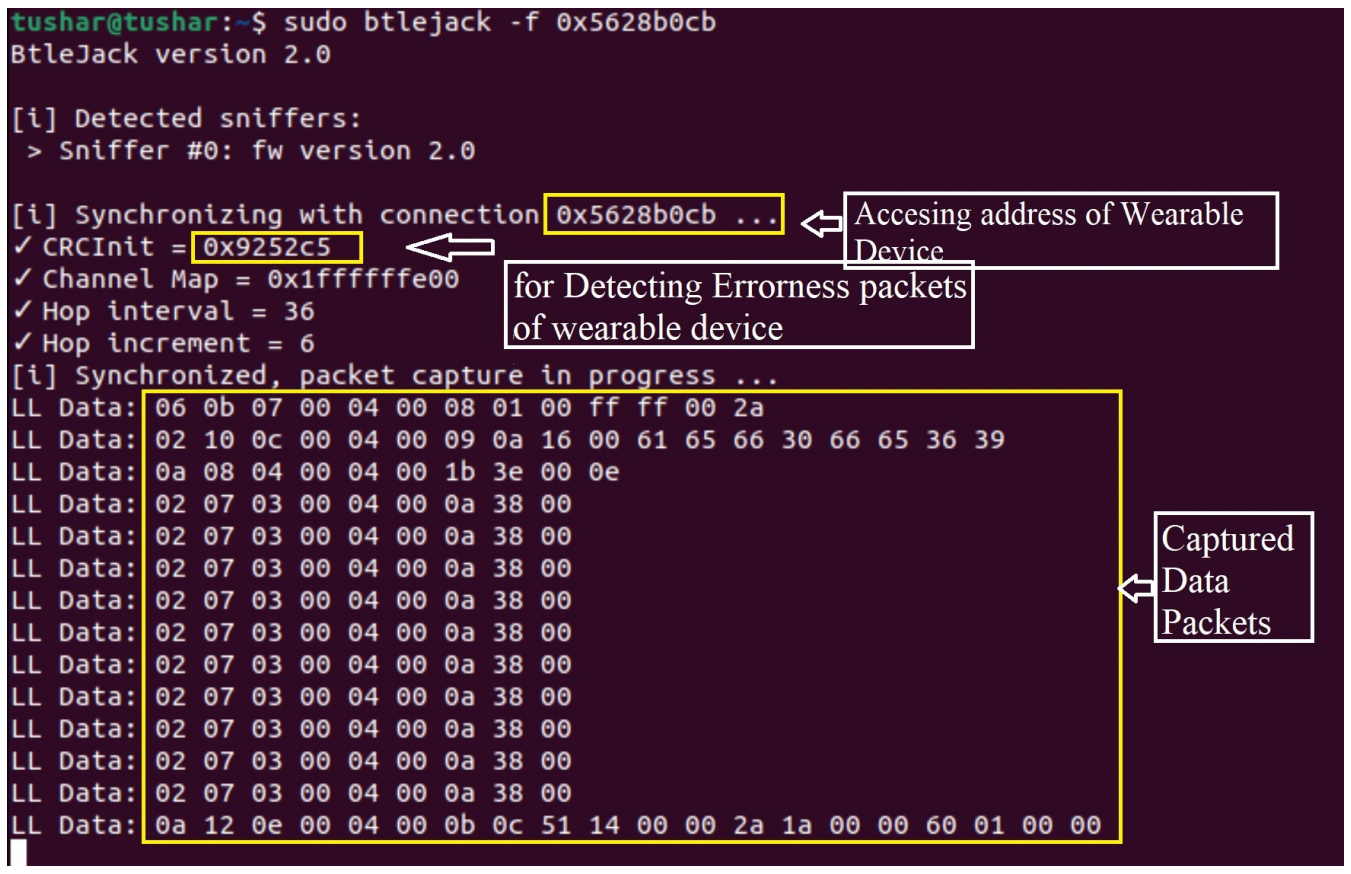

Fig. 9. Sniffing packets using Micro bit V2

## B. Active Attacks

Listening actively, the attacker disrupts communication and takes information in these assaults. MITM and Replay are two active eavesdropping attack versions to compromise data integrity, attackers actively participate in the communication process in MITM [11]. In contrast to MITM, a replay attack does not compromise the sender or recipient instead, the attacker captures the packet and re-transmits it.

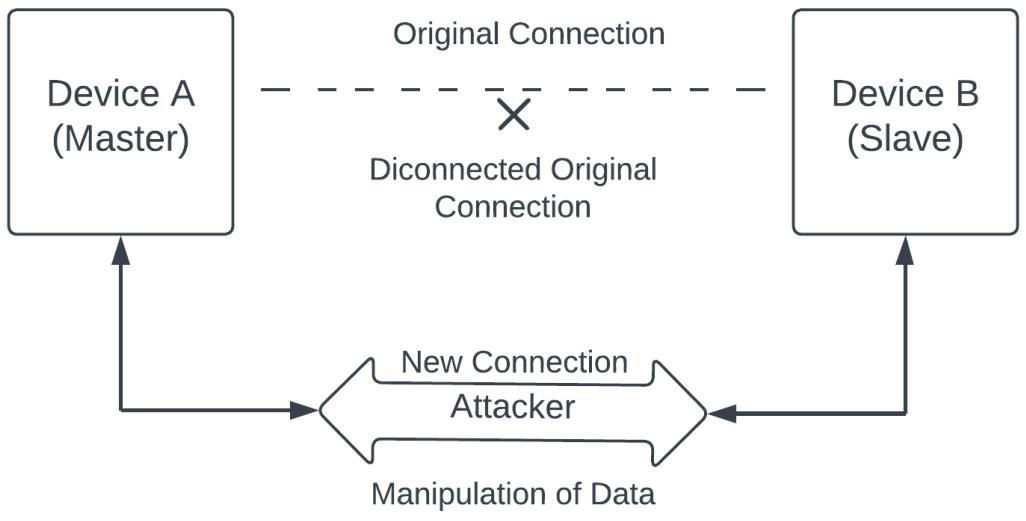

Fig. 10. Active Attacks

*1) Brute-Force BD-ADDR Attack:* As previously stated, the bluetooth protocol employs Frequency Hopping Spread Spectrum (FHSS) to avoid interference by hopping between unique channels of the 2.4GHz ISM radio band [5]. The pseudo-random records utilised to clock this hopping behaviour are obtained from the ”main device” for BD-ADDR to sniff data from a device, you must account for hops, which necessitates knowing the BD-ADDR format. Bluetooth frames do not carry the entire BD-ADDR, but do include the Lower Address Part(LAP), which is a reduction of 24 bits (three bytes). The BD-ADDR is really a combination of the LAP and 1 byte that makes up the upper address part component (UAP). Since the LAPs are broadcast in every frame, the Ubertooth One can capture them passively to calculate the UAP and LAP parts. After a while, the discovery of MAC ID’s acquired from the UAP and LAP sniffing of devices by using Ubertooth-Rx in scan mode (using the -z flag) for a few minutes and letting the gadgets communicate closely with each other.

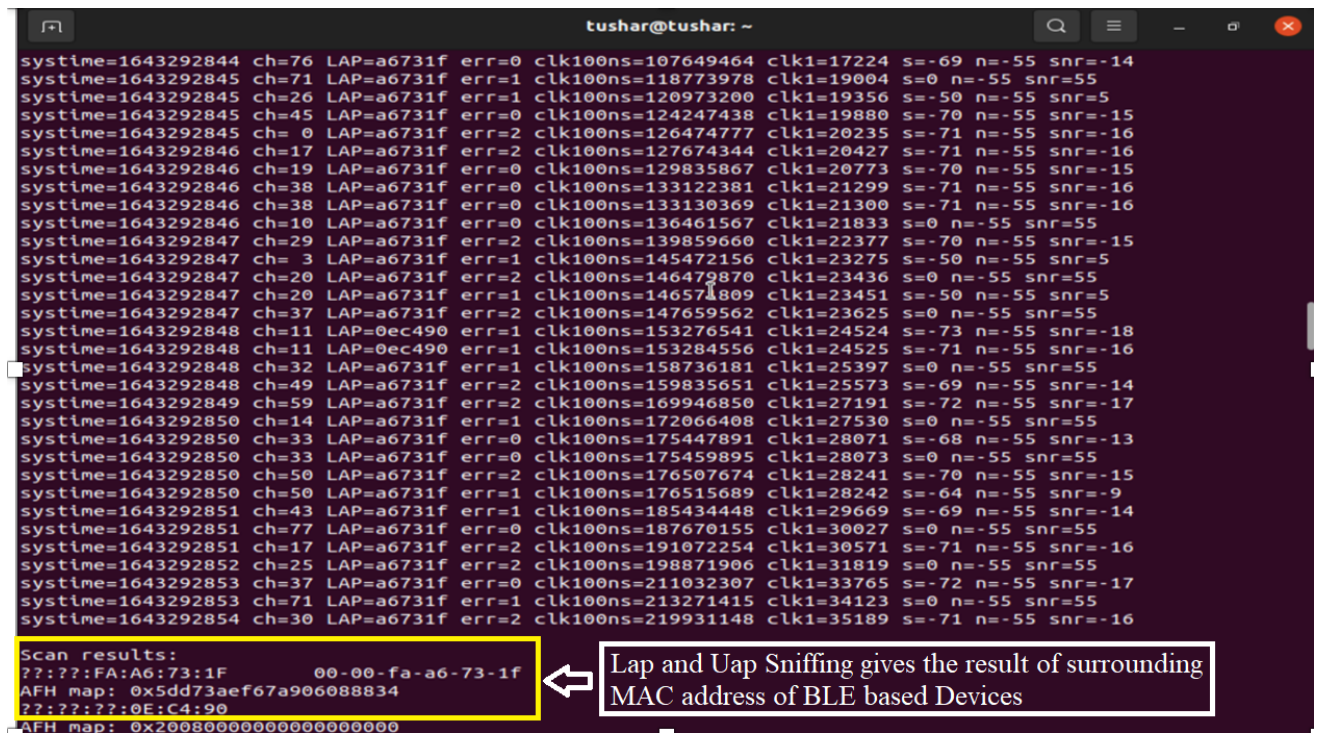

Fig. 11. Sniffing LAP’s and UAP’s using ubertooth to capture MAC Id’s

Ubertooth allows to find gadgets in both concealed and non-discoverable modes. Ubertooth isn’t a full-fledged BTLE device here; it’s just a sniffer that gathers LAP and UAP to form addresses and sends an inquiry to the appropriate BTLE device.

*2) Bluetooth Man-In-The-Middle attack:* There has been researching on MITM attack tactics in IoT systems for both traditional Bluetooth and BLE, and MITM is a common type of wireless communication threat [6]. By placing oneself in the center of the BLE peripheral devices, an attacker may conduct a MITM attack upon them. The attacker intercepts and alters a packet supplied by several devices before being sent to the other. Neither device knows its data is blocked or modified by a suspicious device.

BLE device makers must rigorously adhere to BLE protocol binding and encryption requirements to prevent MITM. It is also best to bypass using the matching approach without updating devices. Secure connections are also recommended for developers since they offer far more robust cryptographic protection than older connections. If the central device (Mobile Phone) is aware that the matching device (MI Band 4) has I/O capabilities, the MITM flag should be supplied during pairing.

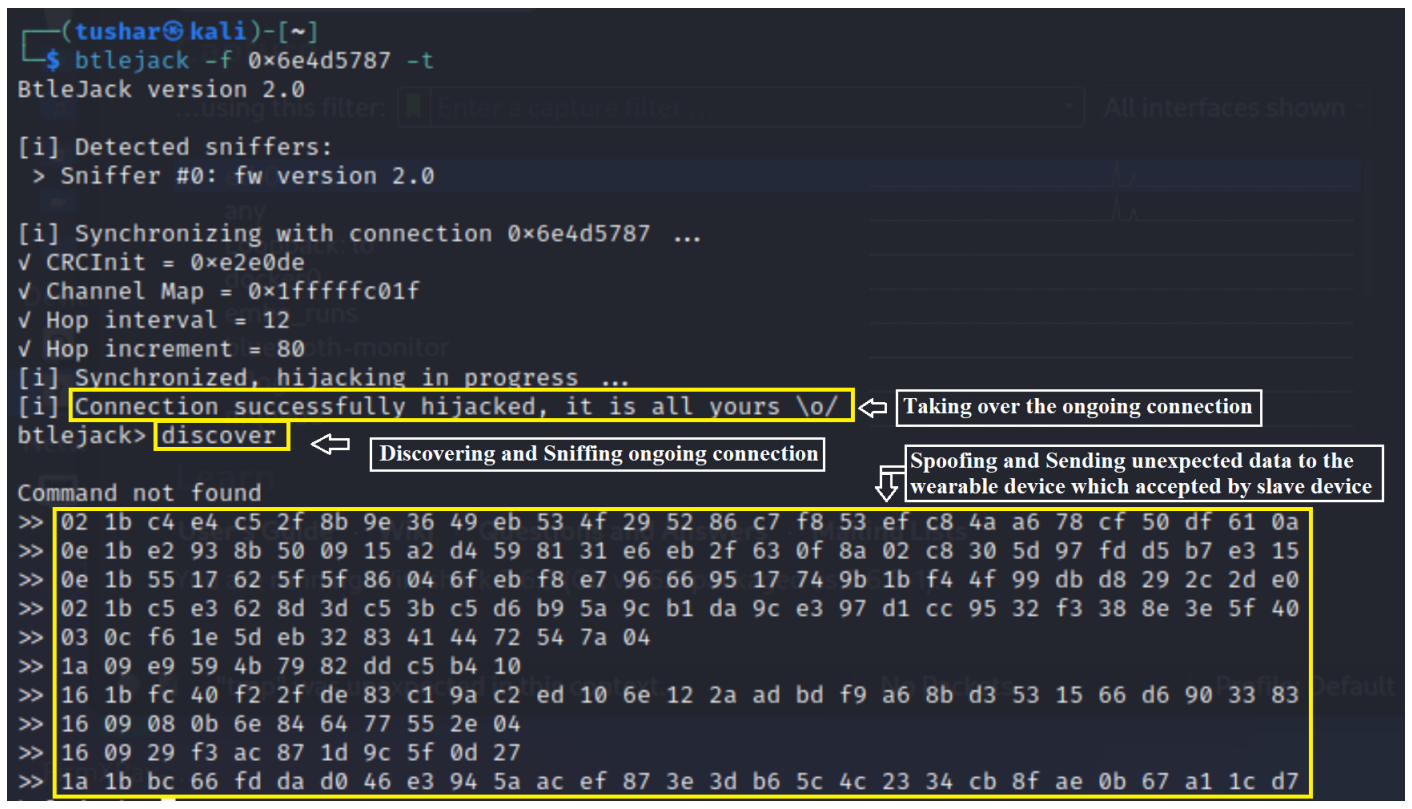

Fig. 12. Man in the Middle Attack

*3) Blue Printing:* Blueprinting gathers precise information on the device's model, manufacturer, unique identifier (IMEI), and software version, with a focus on user privacy issues. This exploit affects both traditional bluetooth and BLE. Although the assault may not do significant damage, it is used to organise future attacks on the victim's device. Blueprinting is not a serious assault, but it does expose personal information. According to the BLE standard, IoT-BLE devices must publicly broadcast their GATT services so that an attacker can obtain this data [17]. Furthermore, attackers can get data on the quantity of devices deployed by a specific manufacturer. If a gadget has a well-known security flaw, this attack can be quite damaging. To carry out this assault, there are several open-source tools accessible. These sources might gather data about the bluetooth stack. Another powerful application for executing this attack quickly is nRF Connect for smartphones, which can be write services directly to the IoT-BLE devices.

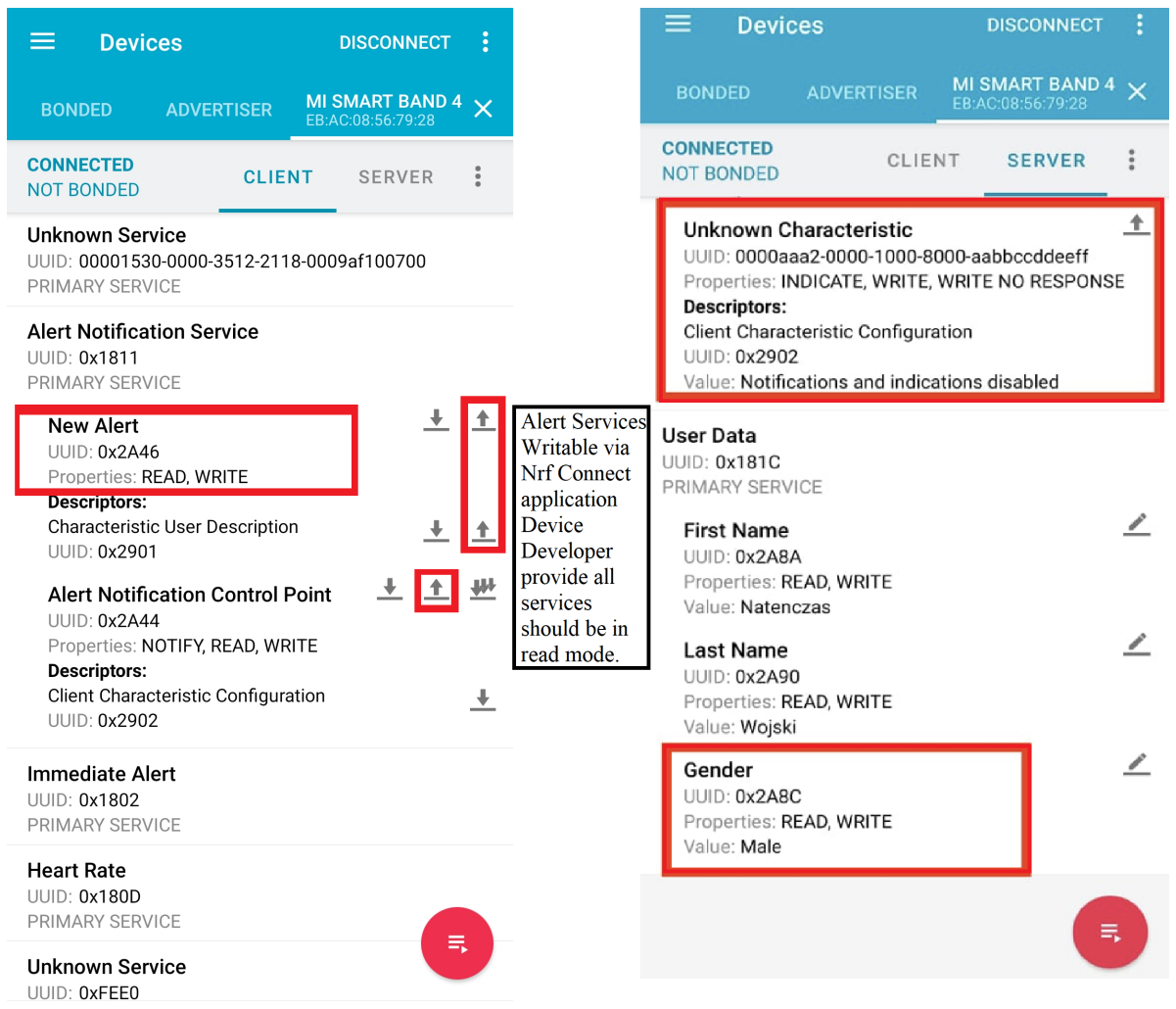

Fig. 13. nRF Connect application

*4) Device Fingerprinting:* Device fingerprinting is used to identify a specific device by utilizing data exclusive to that device, such as the MAC address, UUID, advertising packets, and GATT services. Device fingerprinting infringes user privacy, as seen in fig. 12 above. The fixed MAC addresses of many BLE devices can be used to identify them. By analyzing IoT mobile applications and identifying static UUIDs from ads, several BLE based IoT and Wearable Devices are susceptible to fingerprinting [17].

- UUIDs have a hierarchical structure and need to analyze the set of values to determine the UUID hierarchy for fingerprinting. Simultaneously, it pinpoints application-level flaws such as inappropriate encryption usage.
- Sniffed advertising UUIDs leads to the fingerprinting of IoT devices [13]. Therefore, a single UUID may be utilized by several apps. Connecting to the gadget is indeed required.
- The value set predictable outcome implementation level weaknesses, which leads to the identification of devices that seem to be susceptible to sniffer or unauthorised access.

We need to fix an app-level vulnerability to stop this attack. Developers must provide cryptographically secure features and encrypt credentials. To disguise the UUIDs, implementing use of encryption techniques since the broadcast signal must have a channel level measurement [13]. As a result, the attacker can only receive the annoying signal.

*5) Blue Stumbling:* Finding devices with known security issues is a practice known as blue stumbling. It is a set-up for later, more significant attacks rather than active ones. Using tiny sniffers like the Microbit V2, an attacker sniffs for susceptible devices in a crowded area to remain undetected. The attacker's target devices with security weaknesses.

When a BLE connection is not required, the device must be kept invisible or in un-discoverable mode to prevent attackers from discovering it. Unauthenticated devices must provide just the most basic information.

*6) DOS attack:* Bluetooth-enabled peers can request and receive echoes using the L2CAP protocol in a DoS attack. DoS attack is possible with L2CAPping. This L2CAPping allows seeing the established connection and the round trip time with other Bluetooth-enabled devices. Attacks against smartphones can maintain a minimum of about 10 meters. Powerful transmissions for laptops can reach up to 100 meters. Using standard tools like l2ping, which comes with the Linux Bluex utils package. Several instructions in the l2ping program allow hackers to specify the packet length. Hackers infect Bluetooth-enabled gadgets with malicious programs, rendering them useless to their users. The assault can potentially disrupt the victim's device's regular operation and possibly damage its functionality. The -s number option in the standard BlueZ utility distribution's l2ping allows users to select the packet length for l2ping. Many devices use a packet size of 600 bytes. Creates a packet of the requested size and transmits it to the provided MAC address. As a result, the end device's reaction time grows longer and longer, and the attacked device's Bluetooth capability stops working.

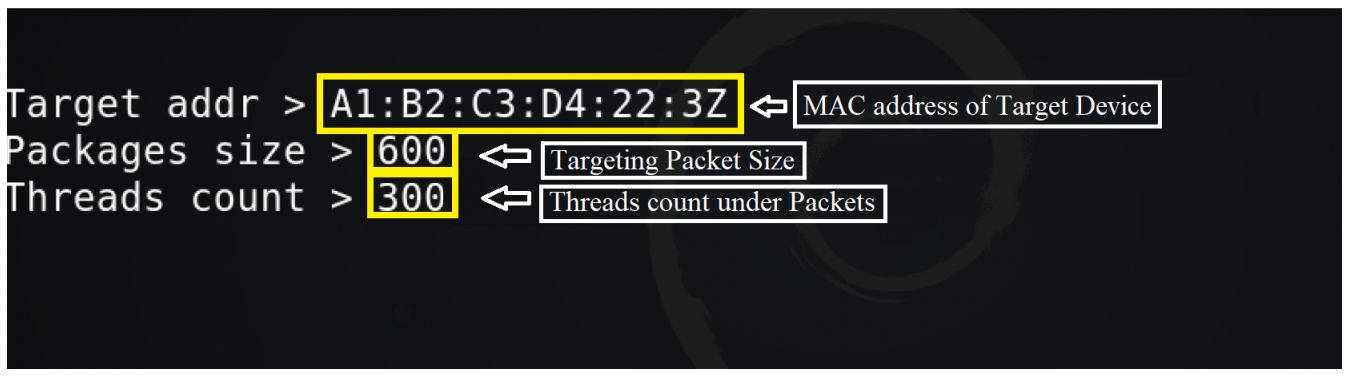

Fig. 14. DOS attack using Adafruit

## V. Summary of BLE Attacks

IoT devices and wearable gadgets that use the BLE protocol have a variety of vulnerabilities. However, lousy protocol design is to blame for many issues. Some BLE security policy devices are subject to cryptographic attack pairing techniques and BLE privacy vulnerabilities. The National Institute of Standards and Technology (NIST) and other security systems researchers have provided specific devices with considerably older versions of (BLE v4.0 and v4.1). As per the authors analysis according to shown in fig. 15 mind map threats to IoT-BLE across the three OSI levels (e.g., physical layer, data link layer, and application layer). Physical layer attacks accompany assaults in the radio spectrum. The bluetooth data connection layer is where the majority of assaults happen. Data transfer via data transfer layer, The attacker captures the link-layer packet and can send the malicious packet via link layer hence link-layer security is a significant problem for BLE developers. Some vulnerabilities are characterized as incorrectly implemented application-layer vulnerabilities by the device maker or developer.

## VI. Future Scope

BLE-enabled smart wearables and IoT gadgets have become a part of our everyday life. Data IoT device management and access control will need to be highly secure. Also, connect to IoT devices via the server rather than users. The gadget has enhanced device management and user privacy dramatically. This server is designed to prevent IoT devices from gaining access to personal data. However, if the network/server is compromised, all devices linked to it are in danger. So there is a chance that researchers may start looking into employing distributed type Blockchain technology to safeguard linked IoT devices shortly. Industries can benefit from a BLE mesh network.

## VII. Conclusion

This paper investigates the effectiveness of IoT-BLE devices. Creating a threat model using the STRIDE threat modelling tool by analyzing and experimenting with IoT-BLE wearable device's attack surface and finding threats is easy to evaluate. Sometimes, its severe only vulnerabilities and different exploitation scenarios like denial of service (DoS), elevation of privilege, information disclosure, spoofing, tampering, and repudiation during the pairing and writable services phase permanently secure this phase with appropriate pairing methods. Moreover, authentication techniques make services UUIDs unreachable for other sources except applications. This paper also gives a quantitative analysis to provide a framework to address security concerns for IoT-BLE wearable devices. BLE is a secure wireless communication protocol, but only if implemented correctly. This investigative analysis with practical proof thoughtfully and practically analyzes the IoT-BLE security procedures and identifying security vulnerabilities during device pairing-related concerns, considering the security measures against the threats and assaults determined after developing specific security measures from threat modeling and the analysis given in this paper. Applying security controls to the system may introduce additional issues, including key management. As a result, it is crucial to repeat the methods described in this work for the security system to assess the security after rebuilding a system.

## Acknowledgment

Authors would acknowledge the support of the Centre of Excellence (CoE) in Complex and Nonlinear Dynamical Systems(CNDS), VJTI, and the members for providing us immense support and a platform for my project.